\documentclass[prb,amsmath,amssymb,superscriptaddress,onecolumn]{revtex4}
\usepackage{float}
\usepackage{amsmath}
\usepackage{amssymb}
\usepackage{amsfonts}
\usepackage{euscript}
\usepackage{enumerate}
\usepackage{hhline}
\usepackage{pslatex}
\usepackage{tabularx}
\usepackage[usenames,dvipsnames]{xcolor}

\usepackage{graphicx}

\usepackage{dcolumn}
\usepackage{bm}
\usepackage[sort&compress]{natbib}

\usepackage{lipsum}

\newcommand{\ks}[1]{\textcolor{black}{#1}}
\newcommand{\xl}[1]{\textcolor{black}{#1}}
\newcommand{\xls}[1]{\textcolor{black}{#1}}
\newcommand{\kss}[1]{\textcolor{black}{#1}}
\newcommand{\xlw}[1]{\textcolor{black}{#1}}
\usepackage[normalem]{ulem}

\begin{document}

\title{Proposal \kss{for} noise-free visible-telecom quantum frequency conversion through \kss{third-order} sum and difference frequency generation}

\author{Xiyuan Lu}\email{xiyuan.lu@nist.gov}
\affiliation{Microsystems and Nanotechnology Division, Physical Measurement Laboratory, National Institute of Standards and Technology, Gaithersburg, MD 20899, USA}
\affiliation{Institute for Research in Electronics and Applied Physics and Maryland NanoCenter, University of Maryland,
College Park, MD 20742, USA}
\author{Gregory Moille}
\affiliation{Microsystems and Nanotechnology Division, Physical Measurement Laboratory, National Institute of Standards and Technology, Gaithersburg, MD 20899, USA}
\affiliation{Joint Quantum Institute, NIST/University of Maryland,
College Park, MD 20742, USA}
\author{Ashutosh Rao}
\affiliation{Microsystems and Nanotechnology Division, Physical Measurement Laboratory, National Institute of Standards and Technology, Gaithersburg, MD 20899, USA}
\affiliation{Institute for Research in Electronics and Applied Physics and Maryland NanoCenter, University of Maryland,
College Park, MD 20742, USA}
\author{Kartik Srinivasan} \email{kartik.srinivasan@nist.gov}
\affiliation{Microsystems and Nanotechnology Division, Physical Measurement Laboratory, National Institute of Standards and Technology, Gaithersburg, MD 20899, USA}
\affiliation{Joint Quantum Institute, NIST/University of Maryland, College Park, MD 20742, USA}

\begin{abstract}
      \noindent Quantum frequency conversion (QFC) between the visible and telecom is a key functionality to connect quantum memories over long distances in fiber-based quantum networks. Current QFC methods for linking such widely-separated frequencies, such as sum/difference frequency generation and four-wave mixing Bragg scattering, are prone to broadband noise from the pump laser(s). To address this issue, we propose to use third-order sum/difference \xlw{frequency} generation (TSFG/TDFG) for an upconversion/downconversion QFC interface. In this process, two pump photons \xls{combine their energy and momentum to} mediate frequency conversion across visible and telecom bands, \xls{bridging a} large spectral gap with long-wavelength pump photons, which is \xls{particularly} beneficial from \xls{the} noise perspective. We show that waveguide-coupled silicon nitride microring resonators can be designed for efficient QFC between 606~nm and 1550~nm via a 1990~nm pump through TSFG/TDFG. We simulate the device dispersion and coupling, and from \xls{the simulated} parameters estimate that the frequency conversion can be efficient ($>80~\%$) at 50~mW pump power. Our results suggest that microresonator-based TSFG/TDFG is promising for compact, scalable, and low power QFC across large spectral gaps.
\end{abstract}

\maketitle

\begin{figure*}[t!]
\centering\includegraphics[width=0.90\linewidth]{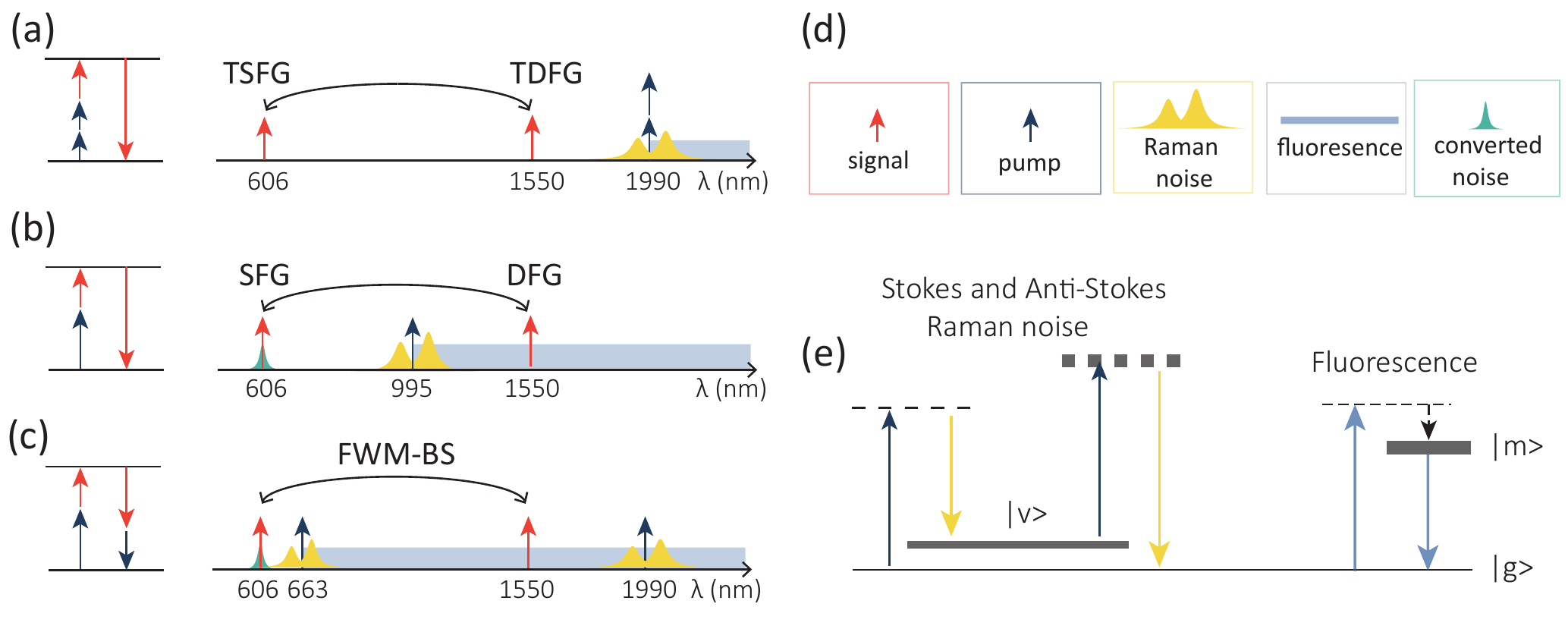}
\caption{\textbf{Proposal \ks{to use third-order} sum/difference frequency generation for visible-telecom QFC.} \textbf{(a)} Energy diagrams (left) and operational schemes (right) of the proposed QFC method using third-order sum/difference frequency generation. The arrow length is scaled by energy/frequency in energy diagrams (left) and normalized in wavelength diagrams (right). \textbf{(b)-(c)} Energy diagrams (left) and operational schemes (right) for current methods for visible-telecom QFC, including (b) $\chi^{(2)}$ sum/difference frequency generation (SFG/DFG) and (c) $\chi^{(3)}$ four-wave mixing Bragg-scattering (FWM-BS).  \textbf{(d)} Besides the signal (red arrow) and pump laser(s) (blue arrow(s)), there are two types of noise processes under consideration, Raman noise (yellow) and fluorescence noise (blue). \kss{Later in the text, we consider the possibility of spontaneous four-wave mixing from the pumps as a third potential noise source.} In consideration of these noise sources, TSFG/TDFG is suitable for above-octave-\kss{spanning} frequency conversion compared to current methods. \xl{\textbf{(e)} Schemes of Stokes and anti-Stokes Raman noise and fluorescence noise, where $|g>$, $|v>$, and $|m>$ represent ground, vibrational, and metastable states, respectively. Dashed lines indicate virtual states. Thicker lines indicate broader spectral lineshapes.}}
\label{Fig1}
\end{figure*}

\noindent Quantum frequency conversion (QFC)~\cite{Kumar_OL_1990} is an important resource to \ks{enable long-distance} interconnects \ks{between visible wavelength quantum systems}, such as optical quantum memories~\cite{Lvovsky_NatPhoton_2009}, \ks{via telecommunications band fiber links in a} quantum network~\cite{Kimble_Nature_2008,Raymer_PhysToday_2012}. \ks{Efficient and low-noise QFC has typically been shown using the $\chi^{(2)}$-mediated process of difference (sum) frequency generation for downconversion (upconversion), in particular in situations where the spectral shift required is small enough that the pump field (whose frequency is equal to the spectral shift) is well-separated from the input signal and frequency-converted idler, and is the longest wavelength involved. This configuration, known to yield low-noise performance in platforms like periodically-poled lithium niobate waveguides~\cite{Pelc_OE_2011}, has been used in a number of demonstrations, for example, to downconvert 910~nm single photons from a quantum dot to 1550~nm~\cite{DeGreve_Nature_2012,Weber_NatNanotech_2019}, corresponding to an $\approx$~140~THz shift.} However, for \xls{quantum memories} operating at shorter wavelengths, \xls{whose} spectral separation \xls{from} the telecom is \xls{more than an octave}, maintaining the pump as the longest wavelength within a single-stage $\chi^{(2)}$ process is no longer feasible. While QFC to the telecom has been shown in experiments linking a 606~nm rare-earth-ion quantum memory to 1550~nm~\cite{Maring_Optica_2018}, as well as 637~nm nitrogen vacancy centers in diamond to 1550~nm~\cite{Dreau_PRAppl_2018}, the strong pump field at a spectral location in-between the input signal and output idler (Fig.~\ref{Fig1}(b)) results in noise (e.g., due to Raman scattering~\cite{Pelc_OE_2011}) that is spectrally aligned with the signal/idler and, for example, causes a degradation in antibunched photon statistics~\cite{Dreau_PRAppl_2018}. \xls{A direct approach is to employ} aggressive, narrowband spectral filtering, \xls{but} typically comes with \xls{excessive} insertion loss. Other approaches to circumvent this challenge include downconversion to 1310~nm ~\cite{Zaske_PRL_2012} and implementing a two-stage conversion process in which a long wavelength pump is used at each stage~\cite{Esfandyarpour_OL_2018},
\xls{both circumventing the challenge of \xlw{direct} QFC above an octave}.

\ks{The $\chi^{(3)}$ nonlinearity has also been studied for QFC via the four-wave mixing Bragg scattering process~\cite{McKinstrie_OE_2005,Raymer_PhysToday_2012}. Here, the frequency shift is defined by the difference in frequencies of two applied pumps. FWM-BS with single photon states has been demonstrated in both optical fibers~\cite{McGuinness_PRL_2010}) and integrated nanophotonic resonators~\cite{Singh_Optica_2019}, though the spectral shifts have generally been small ($<$~20~THz). Recently, FWM-BS has been used to realize frequency conversion between the 1550~nm and 900~nm bands~\cite{Li_NatPhoton_2016}, with conversion efficiency exceeding 60~$\%$. However, noise can be a challenge in this scheme, and in general, the large spectral gaps associated with visible-to-telecom conversion necessitate having one of the pumps situated in-between the signal and idler (Fig.~\ref{Fig1}(c)), similar to \xls{the aforementioned discussion in the} $\chi^{(2)}$ case. As a result, \kss{for visible-telecom QFC using either SFG/DFG ($\chi^{(2)}$) or FWM-BS ($\chi^{(3)})$, to date} one always faces the issue of broadband noise from the pump(s) spectrally overlapping with the \xlw{input and/or output signal}.}

\ks{Thus, while there continue to be significant efforts to mitigate noise and use these existing $\chi^{(2)}$ and $\chi^{(3)}$ approaches to bridge large spectral gaps in QFC experiments, one can also consider whether other nonlinear optical processes might be favorable. \xls{For this purpose,} we propose to use third-order sum/difference \xl{frequency} generation (TSFG/TDFG) as an approach for efficient and low-noise QFC between the visible and telecom bands. This process uses two photons from a degenerate infrared pump to make up the spectral gap between the visible and telecom, with a frequency matching equation given by $\omega_\text{v}=\omega_\text{t}~+~2\omega_\text{p}$, where \xls{$\{\omega_\text{v}, \omega_\text{t}, \omega_\text{p}\}$ represent} the visible, telecom, and infrared pump frequencies, respectively. To be concrete, in Fig.~\ref{Fig1}(a) we consider the visible wavelength of 606~nm (e.g., for a Pr$^{3+}$:Y$_2$SiO$_5$ quantum memory~\cite{Hedges_Nature_2010}), a telecom wavelength of 1550~nm, and propose to use a degenerate pump at $\lambda_\text{p}$~=~1990~nm to make up the spectral difference. A crucial feature of this process, similar to second-order SFG/DFG and FWM-BS, is that each frequency-converted idler photon that is created requires corresponding annihilation of an input signal photon. In contrast, degenerately-pumped four-wave mixing, in which two pump photons are annihilated to create signal and idler photons, can be used to connect widely separated wavelengths~\cite{Lu_NatPhys_2019,Lu_NatPhoton_2019}, but is inherently unsuitable for QFC because there is no direct conversion from signal to idler, but rather from pump to idler~\cite{McKinstrie_OE_2005}. Figure~1 also gives a qualitative indication of how this approach can sidestep broadband noise processes associated with the pump. For $\chi^{(3)}$ media, the main noise processes of concern are Raman scattering, fluorescence, and spontaneous four-wave mixing. As both our input signal and output idler are at higher frequencies than the pump, we anticipate a limited impact from Raman noise or fluorescence. This is in contrast to FWM-BS (Fig.~\ref{Fig1}(b,bottom)), where the 663~nm pump may be a source of broadband fluorescence. Finally, as we show later, our pump is also situated in the dispersion regime which limits spectral extent over which spontaneous four-wave mixing occurs \xl{(so that it does not overlap with the output 1550~nm light)}.}

\begin{figure*}[t!]
\centering\includegraphics[width=0.90\linewidth]{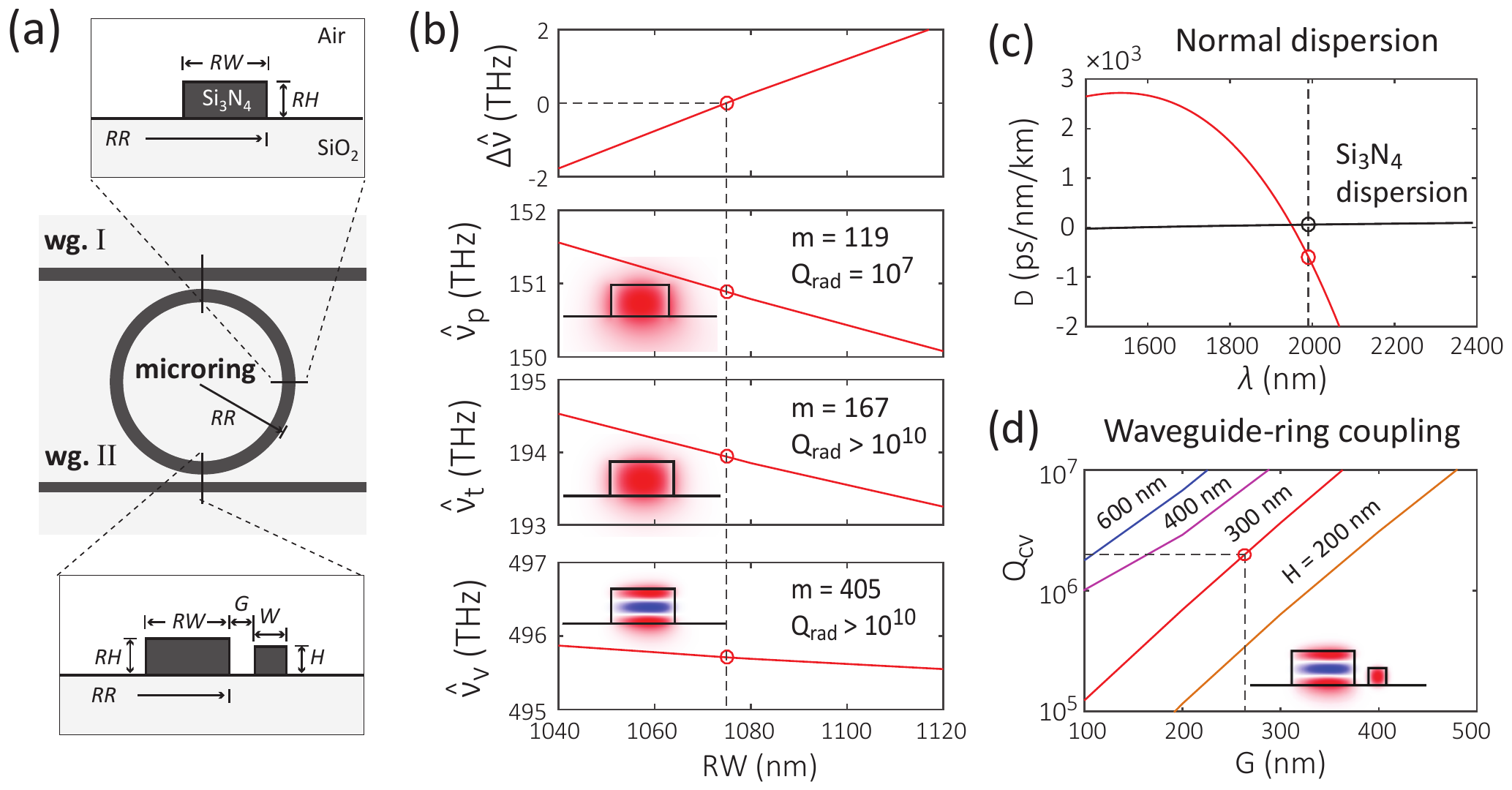}
\caption{\textbf{TSFG/TDFG design in a Si$_3$N$_4$ microring.} \textbf{(a)} Device schematic showing Si$_3$N$_4$ microring coupled with two waveguides. The top microring cross-section determines the dispersion design and the bottom microring-waveguide coupling cross-section determines the coupling design. \textbf{(b)} TE1-TE1-TEv3 dispersion design. The pump and telecom are TE1 modes, and the visible is a TEv3 mode (third-order in vertical direction). The nominal parameters are $RR$ = 25 $\mu$m, $RH$ = 600 nm, and $RW$ = 1075 nm. \textbf{(c)} \xl{Simulations show that the pump mode has normal dispersion ($D = - 580 $~ps/nm/km, \xls{the red circle}) at 1990~nm, while the material dispersion is anomalous ($D = 60$~ps/nm/km, \xls{the black circle}).} \textbf{(d)} Coupling Q simulation for TEv3 visible mode. A waveguide \xls{with $W$ = $H$ = 300~nm and $G~\approx~$270~nm} yield\xls{s} a coupling Q factor ($Q_\text{CV}$) of $2\times10^6$ (dashed lines). Inset shows the coupling geometry.}
\label{Fig2}
\end{figure*}

\ks{While TSFG/TDFG is advantageous with respect to the spectral separation of the pump from the signal and idler, it remains to be seen whether this process can be effectively phase- and frequency-matched in a platform suitable for achieving high efficiency}. \ks{We investigate the TSFG/TDFG process using fully vectorial electromagnetic mode simulations for an integrated silicon nitride (Si$_3$N$_4$) microring resonator, a platform that has been established for $\chi^{(3)}$ nonlinear optical processes~\cite{Moss_NatPhoton_2013}, including those involving widely separated fields, such as third harmonic generation (THG)~\cite{Levy_OE_2011}, FWM-BS~\cite{Li_NatPhoton_2016}, and telecom-visible entangled photon pair generation~\cite{Lu_NatPhys_2019} and classical spectral translation~\cite{Lu_NatPhoton_2019}.}

\xlw{In a microring, assuming perfect frequency matching and zero laser-cavity detuning, the photon flux/number conversion efficiency of TSFG/TDFG is given by (see supplementary material for derivation):
\begin{eqnarray}
\frac{n_\text{out}}{n_\text{in}} = \frac{\Gamma_\text{ct}\Gamma_\text{cv}}{[\Gamma_\text{tt}\Gamma_\text{tv}/(4\gamma U_\text{p})+\gamma U_\text{p}]^2}, \label{eq1}
\end{eqnarray}
\kss{where $n_\text{out}$ and $n_\text{in}$ are the frequency-converted output photon flux and input signal photon flux in the waveguides, respectively. The conversion efficiency is symmetric for telecom-to-visible and visible-to-telecom conversion, that is, $\{$n$_\text{out}$, n$_\text{in}\}$ = $\{$n$_\text{t}$, n$_\text{v}\}$ or $\{$n$_\text{v}$, n$_\text{t}\}$.} $\Gamma_\text{cj}$ and $\Gamma_\text{tj}$ are the coupling and total decay rate for the $j$ mode. $\gamma$ represents the $\chi^{(3)}$ interaction strength of TSFG/TDFG. $U_\text{p}$ is the intra-cavity pump energy, and is related to input pump power by $P_\text{p}=(\Gamma_\text{tp}/2)^2 U_\text{p}/\Gamma_\text{cp}$. The number conversion efficiency is optimized at $\Gamma_\text{ct}\Gamma_\text{ci}/(\Gamma_\text{tt}\Gamma_\text{ti})$ with an intra-cavity pump energy of $U_\text{p} = \sqrt{\Gamma_\text{tt} \Gamma_\text{tv}}/(2\gamma)$.} \xlw{Importantly, $\gamma$ depends solely on the device geometry and its modal phase matching scheme, and is proportional to $\eta$/$V$, where $\eta$ is the mode overlap and $V$ is the averaged mode volume.}

\xlw{Because the frequency span of TSFG/TDFG is similar to that of THG, similar modal phase matching techniques in THG~\cite{Carmon_NatPhys_2007,Levy_OE_2011, Surya_Optica_2018} can be used in this work, that is, using high order modes like TEh5 (transverse-electric mode with five field lobes laterally) or TEv3 (transverse-electric mode with three field lobes vertically) for visible light~\cite{Surya_Optica_2018}}. We here use TEv3 mode at the visible ($\approx$ 606~nm) as an example for TSFG/TDFG, with pump and telecom both in TE1 modes. \ks{The spatial profiles for these modes are shown in Fig.~\ref{Fig2}(b)-(c).} \kss{Another example using TEh5 at visible is described in supplementary, whose performance is not as compelling as the TEv3-based scheme.} The dispersion design for the TEv3 scheme, or more accurately, TE1-TE1-TEv3 for pump-telecom-visible modes, is shown in Fig.~\ref{Fig2}(b). The nominal parameters of $RR$ = 25~$\mu$m, $RH$ = 600~nm, and $RW$ = 1075~nm yield perfect phase matching, \xls{that is, momentum conservation in the azimuthal direction for a micoring.} \ks{In particular, the azimuthal mode numbers $\{m_\text{p}$, $m_\text{t}$, $m_\text{v}\}$ = $\{$119, 167, 405$\}$, satisfy $2m_\text{p}+m_\text{t} = m_\text{v}$, and have resonant frequencies near the targeted values. These modes are confirmed to be frequency matched, with a near-zero $\Delta  \hat{\nu} =  \hat{\nu}_\text{v}- \hat{\nu}_\text{t}-2 \hat{\nu}_\text{p}$ (dashed line).} In the nominal design (circles), the pump, telecom, visible modes are at  150.89 THz (1988.23 nm), 193.94 THz (1546.90 nm), and 495.71 THz (605.19 nm), respectively. The simulated radiation-limited optical quality factors (Q$_\text{rad}$) are $\approx$ 10$^7$ for the pump mode and $>$ 10$^{10}$ for the visible and telecom modes, so that sidewall scattering will likely be the major limitation for the optical quality factors of these modes in practice. Their averaged mode volume ($\bar{V} = (V^2_\text{p}V_\text{t}V_\text{v})^{1/4}$) is calculated to be 61.0~$\mu$m$^3$ and the mode overlap is 8.5~\%. Importantly, the dispersion is normal around the pump, as shown in Fig.~\ref{Fig2}(c), which is beneficial in suppressing optical parametric oscillation~\cite{Vahala_PRL_2004} and frequency comb generation~\cite{Kippenberg_NatPhoton_2019}, as both processes generally require anomalous dispersion. \kss{Their elimination ensures that the pump will be efficiently used for mediating the desired TDFG/TSFG process. It does not necessarily preclude the possibility of spontaneous four-wave mixing (SFWM) from the pump, which can be a noise source if it spectrally overlaps with the converted telecom idler. From energy conservation, this hypothetical SFWM noise process would convert two pump photons at 1988.23~nm to a signal photon at 1546.90~nm and idler photon at 2781.91~nm. For our geometry, however, this process is inhibited, as the ring does not support modes above $\approx$~\xls{2300}~nm, due to cut-off associated with the asymmetric cladding structure.}

We choose a top air cladding and bottom SiO$_2$ substrate for the Si$_3$N$_4$ microring (Fig.~\ref{Fig2}(a)). Such an asymmetric configuration has advantages for both \xls{coupling and dispersion}. \xls{For coupling, long wavelength modes can be cut-off in a narrow waveguide because of this asymmetry, which is necessary to separate the coupling tasks for the widely separated visible and telecom/2~$\mu$m modes~\cite{Lu_NatPhys_2019}.} For dispersion, the ring radius ($RR$), ring width ($RW$), and ring thickness/height ($RH$) are the three geometric control parameters, as indicated in the top cross-section in Fig.~\ref{Fig2}(a). \xls{In air-clad device, these parameters can be trimmed by \xlw{dry/wet etching for dispersion tuning}~\cite{Lu_NatPhoton_2019}.}

\ks{While high-$Q$, relatively small mode volume, and phase- and frequency-matching ensure efficient intra-cavity conversion, for an overall high on-chip efficiency, efficient resonator-waveguide coupling is needed (see Equation~(\ref{eq1})). The waveguide width ($W$), waveguide height ($H$), and ring-waveguide gap ($G$) are three adjustable parameters that, together with the microring geometry, determine the coupling characteristics. A cross-section schematic of the coupling region is shown in  Fig.~\ref{Fig2}(a), and we use a coupled mode theory formalism~\cite{Moille_OL_2019} to determine the coupling quality factor ($Q_\text{c}$) at the three targeted wavelengths, where in all cases the waveguide mode considered is the fundamental TE1 mode. The resonator's TE1 pump ($\approx$ 1990~nm) and telecom modes are easily coupled to the waveguide. \xl{For example, a waveguide with $W$ = 750 nm and $G$ = 650~nm results in $Q_\text{c}$ $\approx 2\times10^6$}. At such a gap, coupling of the visible signal is negligible \xl{($Q_\text{cv} > 10^9$)}. We thus use a second waveguide exclusively for coupling of the TEv3 visible mode in the resonator, where its size (width and height) is small enough that, given the asymmetric cladding, the telecom and pump wavelengths are cut-off. As shown in Fig.~\ref{Fig2}(d), a square waveguide with $W = H$ = 300~nm and $G$ = 270~nm, which can provide $Q_\text{cv} = 2\times10^6$ (red circle). $Q_\text{cv} = 2\times10^5$ can be achieved by either a closer gap or a shallower waveguide (Fig.~\ref{Fig2}(f)).} To implement such waveguide coupling in practice, a second alignment in the electron beam lithography and a separate shallow-etching process \xlw{can be used}.

\begin{figure}[t!]
\centering\includegraphics[width=0.60\linewidth]{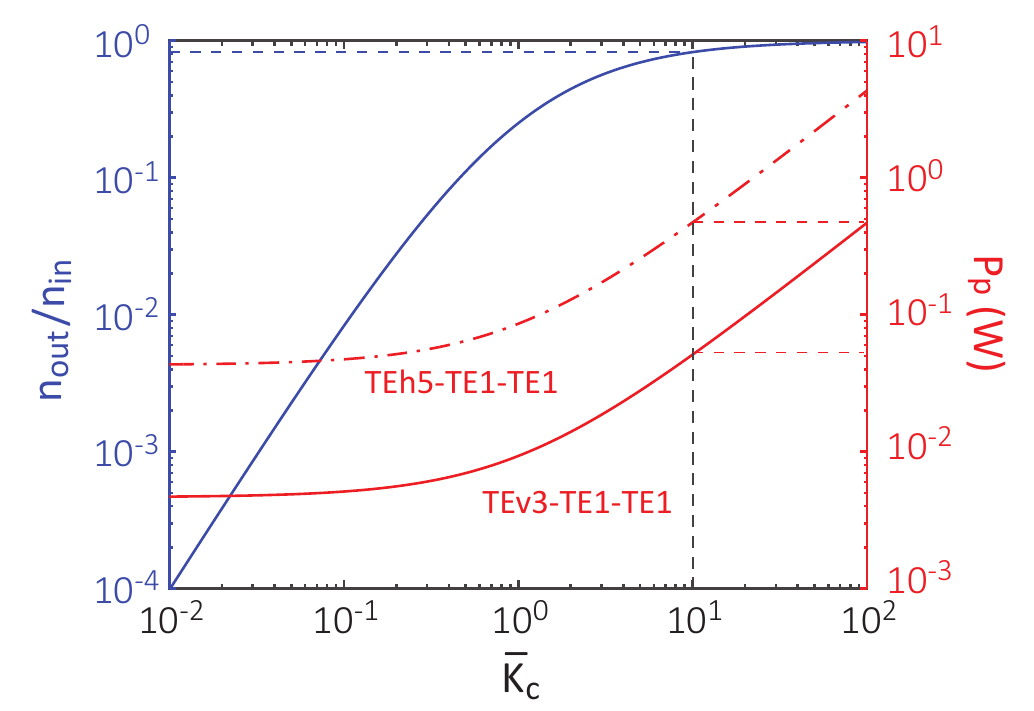}
\caption{\textbf{Optimized conversion efficiency and required pump powers.} In-waveguide photon flux/number conversion efficiency (\xls{blue}, left y-axis) as a function of effective coupling ratio ($\bar{K}_\text{c} \equiv \sqrt{\Gamma_\text{ct}\Gamma_\text{cv}/(\Gamma_\text{0t}\Gamma_\text{0v})}$). \kss{This efficiency is independent of the scheme (TEv3/TEh5 for the visible mode) and the direction of the frequency conversion (visible-to-telecom, or telecom-to-visible). For example, to achieve a number conversion efficiency of 83~\% \xls{(the blue dashed line)}, a $\bar{K}_\text{c}$ of 10 is needed \xls{(the black dashed line)}. However, to achieve such efficiency and effective coupling ratio, the power required (\xls{red}, right y-axis) to operate the TEv3/TEh5 (solid/dot-dashed lines) schemes differs by almost an order of magnitude, \xlw{requiring 50~mW/470~mW \xls{(the red dashed lines)}, respectively.}}}
\label{Fig3}
\end{figure}

Finally, the optimized efficiencies and required pump powers following Equation~(\ref{eq1}) are shown in Fig.~\ref{Fig3}, where we assume all optical modes have intrinsic $Q=2\times10^6$, and consider the case where the pump is critically coupled. For the TEv3 design, at 50 mW input power, with an over-coupling of 10$\times$, the number conversion efficiency is 83~\% (dashed lines in Fig.~\ref{Fig3}). \kss{As noted earlier, stronger overcoupling (either by higher intrinsic $Q$ or lower $Q_\text{c}$) will \xlw{further} increase the conversion efficiency, though the former is preferred as the latter comes at the expense of additional pump power.}

In summary, we propose using third\xlw{-order} sum/difference frequency generation (TSFG/TDFG) for noise-free QFC, and provide detailed simulations for realizing such a process in a Si$_3$N$_4$ microring. This nanophotonic QFC interface could be particularly useful for building scalable quantum networks. Although we only simulate this process in a microring, TSFG/TDFG can be applied to other platforms, including nanophotonic waveguides and periodically-poled crystals.
\newpage

\centering{\textbf{Supplementary Information: Theory and Additional Simulation Data}}

\section{Theory for cavity-enhanced TSFG/TDFG}
\noindent In this section, we give more details on the equations we provide in the main text, starting with the following coupled mode equations~\cite{Lin_OE_2007,Lu_Optica_2019}:
\begin{eqnarray}
\frac{d\tilde{A}_\text{p}}{dt} &=& {[i(\Delta\omega_\text{p} +\gamma_\text{pppp} U_\text{p}) - \Gamma_\text{tp}/2]}~\tilde{A}_\text{p}+ 2i \gamma_\text{ptpv} \tilde{A}_\text{s} \tilde{A}^*_\text{t} \tilde{A}^*_\text{p} + i \sqrt{\Gamma_\text{cp}} \tilde{S}_\text{p}, \label{EqS1}\\
\frac{d\tilde{A}_\text{v}}{dt} &=& {[i(\Delta\omega_\text{v} + 2\gamma_\text{vpvp} U_\text{p}) - \Gamma_\text{tv}/2]}~\tilde{A}_\text{v} + i \gamma_\text{vptp} \tilde{A}^2_\text{p} \tilde{A}_\text{t} + i \sqrt{\Gamma_\text{cv}} \tilde{S}_\text{v}, \label{EqS2}\\
\frac{d\tilde{A}_\text{t}}{dt} &=& {[i(\Delta\omega_\text{t} + 2\gamma_\text{tptp} U_\text{p}) - \Gamma_\text{tt}/2]}~\tilde{A}_\text{t} + i \gamma_\text{tpvp} (\tilde{A}^*_\text{p})^2 \tilde{A}_\text{v}+i \sqrt{\Gamma_\text{ct}} \tilde{S}_\text{t}.  \label{EqSs}
\end{eqnarray}
As mentioned in the text, the equations already assume perfect phase matching, that is, $m_\text{v} = 2 m_\text{p}+m_\text{t}$. The field amplitude is normalized so that $U_\text{i} = |\tilde{A}_\text{i}|^2$ where $i$ = $\{p$, $v$, $t\}$ represents intra-cavity energy for pump, signal and idler, respectively. The first terms describe the cavity evolution considering Kerr shifts. The cavity detuning without Kerr shifts is $\Delta \omega_\text{i} = \omega_\text{i} - \omega_\text{i0}$, where $\omega_\text{i0}$ represents the center of the Lorentzian resonances for the $i$ mode. The self/cross phase modulation (SPM/XPM) red-shift/decrease cavity resonance frequencies, depending on pump intra-cavity optical energies ($U_\text{p}$) only, as $U_\text{v}$ and $U_\text{t}$ are at the quantum level for quantum frequency conversion. $\Gamma_\text{ti}$ describes the decay of the intra-cavity energy $U_\text{i}$, which includes the intrinsic cavity loss and the out-coupling to waveguide, $\Gamma_\text{ti} = \Gamma_\text{0i} +\Gamma_\text{ci}$. Here the decay term $\Gamma_\text{ji}$ is related to optical quality factor $Q_\text{ji}$ or the field coupling/decay time $\tau_\text{ji}$ by:
\begin{eqnarray}
\Gamma_\text{ji} = \frac{\omega_\text{0i}}{Q_\text{ji}} = \frac{2}{\tau_\text{ji}},~(j = t,0,c;~i = p,v,t).
\end{eqnarray}
The second terms describe the TSFG/TDFG interaction with degenerate pump, signal and idler. Both Kerr shifts and TSFG/TDFG interaction are $\chi^{(3)}$ processes, and the involved nonlinear interaction term $\gamma_\text{ijkl}$ is given by:
\begin{eqnarray}
\gamma_\text{ijkl} = \frac{3 \omega_\text{i} \eta_\text{ijkl} \chi^{(3)}_\text{ijkl}}{4 \bar{n}^4_\text{ijkl} \epsilon_\text{0} \bar{V}_\text{ijkl}},~(\text{with~i,j,k,l = p,v,t}),
\end{eqnarray}
which is a positive real parameter. For simplicity, we here assume the ideal situation of perfect frequency matching after the Kerr effect (the Kerr-shifted cavity detuning is zero). Therefore, we only need to consider the TSFG/TDFG interaction, and we have:
\begin{eqnarray}
\frac{\gamma}{\sqrt{\omega_\text{v}\omega_\text{t}}} \equiv \frac{\gamma_\text{pvpt}}{\omega_\text{p}} = \frac{\gamma_\text{vptp}}{\omega_\text{v}} = \frac{\gamma_\text{tpvp}}{\omega_\text{t}} = \frac{3 \eta_\text{pptv} \chi^{(3)}_\text{pptv}}{4 \bar{n}^4_\text{pptv} \epsilon_\text{0} \bar{V}_\text{pptv}}, \label{Eq_gamma}
\end{eqnarray}
where $\gamma$ is used in the main text so that it is independent of telecom or visible modes. The permutation sequence in the last term does not matter. Therefore, we abbreviate the subscripts of $\bar{n}$, $\eta$, $\bar{V}$, and $\chi^{(3)}$ in the main text. Both $\eta$ and $\bar{V}$ are calculated from the pump, telecom, and visible mode profiles,
\begin{eqnarray}
\eta &=& \frac{\int_\text{V}dv~\epsilon_\text{p}\sqrt{\epsilon_\text{v}\epsilon_\text{t}} \tilde{E}^2_\text{p} \tilde{E}_\text{t}\tilde{E}^*_\text{v}} {(\int_\text{V}dv~\epsilon^2_\text{p} {|\tilde{E}_\text{p}|}^4)^{1/2}
(\int_\text{V}dv~\epsilon^2_\text{v} {|\tilde{E}_\text{v}|}^4 \int_\text{V}dv~\epsilon^2_\text{t} {|\tilde{E}_\text{t}|}^4)^{1/4}}, \label{Eq_eta}\\
\bar{V} &=& (V^2_\text{p}V_\text{t}V_\text{t})^{1/4}, ~\text{where}~V_\text{i} = \frac{({\int_\text{V} dv~\epsilon_\text{i}|\tilde{E}_\text{i}|}^2)^2}{\int_\text{V}dv~\epsilon^2_\text{i} {|\tilde{E}_\text{i}|}^4}~\text{(with i = p,t,v).} \label{EqV}
\end{eqnarray}
The last terms in these two equations are the source terms for pump and signal, where $P_\text{i}=|\tilde{S}_\text{i}|^2$ represents the input power in the waveguide.

In perfect cavity detuning and steady state, considering the visible and telecom fields are at the quantum level, and only considering upconversion (where the visible mode has no input), the equations of the cavity fields are reduced to:
\begin{eqnarray}
 (\Gamma_\text{tp}/2)^2U_\text{p} &=& \Gamma_\text{cp} P_\text{p} ~\label{Eq_Pp}\\
 (\Gamma_\text{tv}/2)^2U_\text{v} &=& \gamma_\text{vptp}^2 U^2_\text{p} U_\text{t} ~\label{Eq_Psi}\\
 (\Gamma_\text{tt}/2+2\gamma_\text{tpvp}^2  U^2_\text{p}/\Gamma_\text{tv})^2 U_\text{t} &=& \Gamma_\text{ct} P_\text{t} ~\label{Eq_ratio}
\end{eqnarray}
The visible output power is the out coupling of the cavity energy, that is:
\begin{eqnarray}
P_\text{v} = \Gamma_\text{cv} U_\text{v}.~\label{Eq_Pi}
\end{eqnarray}
In combination of Eq.~(\ref{Eq_Psi}) and Eq.~(\ref{Eq_gamma}), we have the intra-cavity photon number ratio as,
\begin{eqnarray}
\frac{N_\text{v}}{N_\text{t}} = \frac{\omega_\text{t}}{\omega_\text{v}} \frac{U_\text{v}}{U_\text{t}} = (\frac{\gamma U_\text{p}}{\Gamma_\text{tv}/2})^2.
\end{eqnarray}
\xls{In the waveguide,} considering this equation along with Eq.~(\ref{Eq_ratio}) and Eq.~(\ref{Eq_Pi}), we have the number/flux conversion efficiency as:
\begin{eqnarray}
\frac{n_\text{v}}{n_\text{t}} = \frac{\omega_\text{t}}{\omega_\text{v}} \frac{P_\text{v}}{P_\text{t}} = \frac{\Gamma_\text{ct}\Gamma_\text{cv}}{[(\Gamma_\text{tt}\Gamma_\text{tv}/(4\gamma U_\text{p})+\gamma U_\text{p}]^2}. \label{Eq_solution}
\end{eqnarray}
We can see from Eq.~(\ref{Eq_solution}) that the number conversion efficiency is optimized with a value of $\Gamma_\text{ct}\Gamma_\text{ci}/(\Gamma_\text{tt}\Gamma_\text{ti})$ when the intracavity pump energy satisfies $\gamma U_\text{p} = \sqrt{\Gamma_\text{tt} \Gamma_\text{tv}}/2$, with pump input power given by Eq.~(\ref{Eq_Pp}). The number conversion efficiency approaches unity (100~\%) when both visible and telecom modes are significantly overcoupled ($\Gamma_\text{tt} \approx \Gamma_\text{ct}$, $\Gamma_\text{tv} \approx \Gamma_\text{cv}$). Although we have only consider the upconversion case explicitly, the expressions in terms of number ratio/efficiency are the same for downconversion in Eq.~\ref{Eq_solution}. Therefore, in the main text, we use  $\{$n$_\text{out}$, n$_\text{in}\}$ to represent both cases, that is, $\{$n$_\text{t}$, n$_\text{v}\}$ for downconversion and or $\{$n$_\text{v}$, n$_\text{t}\}$ for upconversion.

\begin{figure*}[t!]
\centering\includegraphics[width=1.0\linewidth]{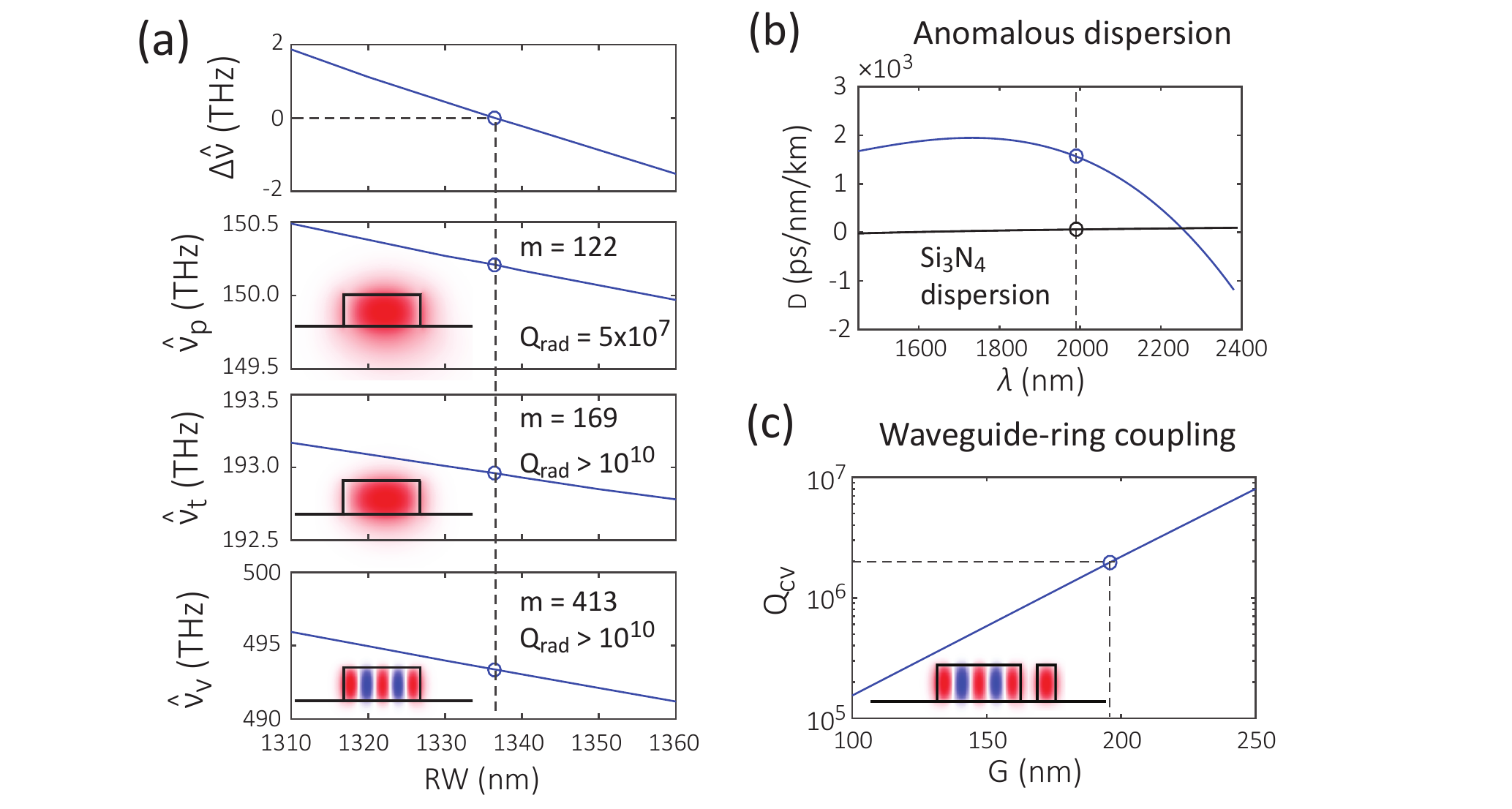}
\caption{\textbf{An alternative TSFG/TDFG design.} \textbf{(a)} TE1-TE1-TEh5 dispersion design. The mode profiles (radial component of electric field) are shown in the insets, \xl{with radiation Q  values ($Q_\text{rad}$) specified for pump, telecom, and visible modes}. The pump and telecom are fundamental transverse-electric-like (TE1) modes, while the visible is a TEh5 mode. The nominal parameters are $RR$ = 25 $\mu$m, $RH$ = 600 nm, and $RW$ = 1337 nm, which leads to perfect frequency matching (circles). \textbf{(b)} \xl{Simulation shows that this design exhibits anomalous dispersion around the pump mode at 1990~nm ($D = 1,600$~ps/nm/km, \xls{the blue circle}), and the material dispersion is also anomalous ($D = 60$~ps/nm/km, \xls{the black circle}).} \xl{This dispersion corresponds to 3.8~GHz (anomalous) in frequency mismatch of the adjacent modes.} \textbf{(c)} Coupling Q simulation for TEh5 visible mode. The waveguide with $RW$ = 300~nm, $RH$ = 600~nm, and $G~\approx~$200~nm has a coupling Q factor ($Q_\text{CV}$) of $2\times10^6$ (dashed lines). Inset shows its microring-waveguide coupling geometry.}
\label{FigS2}
\end{figure*}
\section{Additional simulation data for TEh5-TE1-TE1 configuration}
\noindent \xls{The dispersion design for the TEh5 scheme, or more accurately, TE1-TE1-TEh5 for pump-telecom-visible modes, is shown in Fig.~\ref{FigS2}(a). The nominal parameters of $RR$ = 25~$\mu$m, $RH$ = 600~nm, and $RW$ = 1337~nm yield perfect phase matching, \ks{which for a microring, corresponds to momentum conservation in the azimuthal direction. In particular, the azimuthal mode numbers $\{m_\text{p}$, $m_\text{t}$, $m_\text{v}\}$ = $\{$122, 169, 413$\}$, satisfy $2m_\text{p}+m_\text{t} = m_\text{v}$, and have resonant frequencies near the targeted values. In the nominal design (circles), the pump, telecom, visible modes are at 150.21~THz (1997.20~nm), 192.96~THz (1554.73~nm), and 493.98~THz (608.05~nm), respectively. Their averaged mode volume, given by $\bar{V} = (V^2_\text{p}V_\text{t}V_\text{v})^{1/4}$, is calculated to be 71.8~$\mu$m$^3$ and the mode overlap is 1.1~\%.} In addition, this device geometry shows anomalous dispersion around the pump ($D = 1,600$~ps/nm/km), as shown in  Fig.~\ref{FigS2}(b), which corresponds to 3.8~GHz frequency mismatch in adjacent modes with a 1~THz free spectral range. The larger mode volume and smaller mode overlap, and the anomalous dispersion around the pump make this TEh5 scheme less appealing than the proposed TEv3 scheme. One advantage of the TEh5 scheme is its simplicity in coupling design. As shown in Fig.~\ref{FigS2}(c), a waveguide with $W$ = 300~nm and $G$ $\approx$ 200~nm can provide $Q_\text{cv}$ = $2\times10^6$ (dashed lines). Such a coupling is more straightforward then the TEv3 configuration, and does not requires a separate shallow etching process.}

\medskip
\noindent \textbf{Funding.}
This work is supported by the DARPA DODOS and NIST-on-a-chip programs.

\smallskip
\noindent \textbf{Acknowledgements.}
X.L. \xlw{and A.R.} acknowledge support under the Cooperative Research Agreement between the University of Maryland and NIST-PML, Award no. 70NANB10H193.

\bibliography{TSFG_FC.bib}

\end{document}